\documentclass[a4paper,twocolumn,11pt,unpublished]{quantumarticle}

\pdfoutput=1
\usepackage[utf8]{inputenc}
\usepackage[english]{babel}
\usepackage[T1]{fontenc}
\usepackage{amsmath}
\usepackage[unicode=true, breaklinks=false, pdfborder={0 0 1}, backref=false, colorlinks=true, allcolors = blue, linkcolor=blue, citecolor=blue]{hyperref}

\usepackage{tikz}
\usepackage{lipsum}

\usepackage[numbers,sort&compress]{natbib}

\usepackage{amssymb}
\usepackage{amsthm}
\usepackage{amsfonts}
\usepackage{graphicx}
\usepackage{enumitem}
\usepackage{bm}
\usepackage{mathtools}
\usepackage{leftindex}
\usepackage{rotating}
\usepackage{pbox}
\usepackage[dvipsnames]{xcolor}
\usepackage{array}
\usepackage{physics}
\usepackage{dsfont}
\usepackage{mathtools}
\usepackage{leftidx}
\usepackage[normalem]{ulem}
\usepackage{braket}
\usepackage{tensor}
\usepackage{mathrsfs}
\usepackage{float}
\usepackage{dsfont}
\usepackage[margin=2cm]{geometry}
\usepackage[title]{appendix}
\usepackage{tcolorbox}
\usepackage{titlesec}

\renewcommand{\bar}{\overline}
\renewcommand{\tilde}{\widetilde}

\newcommand{\rr}[1]{\left(#1\right)}

\newcommand{\EE}{\mathcal{E}}
\newcommand{\CC}{\mathcal{C}}

\begin{document}

\title{Entanglement transference and non-inertial quantum reference frames}

\author{Everett A. Patterson}
\email{ea2patte@uwaterloo.ca}
\affiliation{Department of Physics and Astronomy, University of Waterloo, Waterloo, Ontario, N2L 3G1, Canada}
\affiliation{Institute for Quantum Computing, University of Waterloo, Waterloo, Ontario, N2L 3G1, Canada}
\affiliation{Waterloo Centre for Astrophysics, University of Waterloo, Waterloo, Ontario, N2L 3G1, Canada}
\affiliation{Perimeter Institute for Theoretical Physics, 31 Caroline St N, Waterloo, Ontario, N2L 2Y5, Canada}

\author{Sijia Wang}
\email{s676wang@uwaterloo.ca}
\affiliation{Department of Physics and Astronomy, University of Waterloo, Waterloo, Ontario, N2L 3G1, Canada}
\affiliation{Waterloo Centre for Astrophysics, University of Waterloo, Waterloo, Ontario, N2L 3G1, Canada}

\author{Robert B. Mann}
\email{rbmann@uwaterloo.ca}
\affiliation{Department of Physics and Astronomy, University of Waterloo, Waterloo, Ontario, N2L 3G1, Canada}
\affiliation{Institute for Quantum Computing, University of Waterloo, Waterloo, Ontario, N2L 3G1, Canada}
\affiliation{Waterloo Centre for Astrophysics, University of Waterloo, Waterloo, Ontario, N2L 3G1, Canada}
\affiliation{Perimeter Institute for Theoretical Physics, 31 Caroline St N, Waterloo, Ontario, N2L 2Y5, Canada}


\begin{abstract}

Given the recent interest in perspectival quantum reference frames (QRFs), we ask how quantum properties in the perspectival picture relate to their global (non-perspectival) counterparts. 
Such a connection could allow established quantum information to be understood from the perspective of QRFs.
Specifically, we find sufficient conditions under which global entanglement decomposes into a combination of perspectival entanglement and coherence---a phenomenon that we call \emph{entanglement transference}. We apply this result to non-inertial QRFs, revisiting the problem of acceleration-induced entanglement degradation. 
We find that entanglement degradation in the perspectival picture can be offset by an increase in coherence resources. 
This novel insight into (relativistic) quantum resource transformations suggests that QRFs may play a useful role in understanding more general quantum resource and quantum information phenomena.

\end{abstract}

\maketitle

\section{INTRODUCTION} \label{sec:introduction}

In the last decade, there has been renewed interest in the idea of quantum reference frames (QRFs). These developments have been largely motivated as the formulation of a new toolkit to make sense of quantum gravity, though they are of interest to the broader fields of quantum foundations and quantum information. Many distinct and related notions of QRFs have been proposed in recent years \cite{Hoehn:2019fsy,Castro-Ruiz:2025yvi,DeVuyst:2025ezt}, including the \emph{perspectival} approach to QRFs \cite{Giacomini:2017zju,delaHamette:2020dyi}. Inspired by relational quantum mechanics \cite{rovelli_relational_1996}, this approach stipulates that the quantum properties of a system should be understood from the perspective of its quantum subsystems. It is further assumed that each subsystem to which we have assigned a quantum reference frame, i.e., a perspective, cannot perceive itself to be in a quantum superposition\footnote{Though not explicitly stated this way previously, the assumption that each subsystem perceives itself to be in a state without coherence is quite important to the perspectival approach to QRFs. In the literature, this has previously been described as a ``default zero-state.'' \cite{delaHamette:2020dyi}}.

Despite recent progress, the relationship between perspectival QRFs and standard quantum information theory is still poorly understood. 
In the standard approach to quantum theory, microscopic quantum systems are described from the perspective of an implicit macroscopic observer, with the two separated by a so-called Heisenberg cut \cite{Brukner:2015gcd}.
Most descriptions of quantum information describe the quantum states and observables of interest from such an ``outsider'' view.
In this paper, we refer to this as the \emph{global} or \emph{non-perspectival} picture. By contrast, the perspectival approach is concerned with how quantum systems would describe each other. Since most well-known features of quantum information theory, such as quantum channels, resources, and computing \cite{Nielsen:2012yss}, are derived in the global picture, it is natural to ask how these might be described by, or relate to, a perspectival approach.

In this paper, we find that for sufficiently simple systems the sum of entanglement and coherence of a perspectival state is equal to the entanglement of its associated global state. This connection between the quantum resources of the global state and of the perspectival state can be simply expressed as $\EE_p+\CC_p=\EE_g$, where $\EE$ and $\CC$ denote the entanglement and coherence measures with the subscripts $p$ and $g$ indicating whether the state is perspectival or global. We refer to this phenomena as \emph{entanglement transference} to describe how the global entanglement is transformed into a combination of perspectival entanglement and perspectival coherence. We specifically find that for a 3-qubit system, entanglement transference is guaranteed across all perspectives for global states that are either \emph{even} or \emph{odd}, i.e., comprised of computational basis states with either an even or odd number of $\ket{1}$'s.

In order to allow us to relate these perspectival and global states in the manners above, we construct an explicit perspective assignment procedure.
With this tool in hand, we are then able examine relativistic quantum information (RQI) problems from the perspective of QRFs.
In particular, we revisit the problem of entanglement degradation in RQI, providing a novel explanation in terms of perspectival QRFs and entanglement transference.

The discovery that relative accelerated motion between two observers Alice and Rob leads to a degradation in the entanglement between a pair of quantum field modes
in (classical) non-inertial reference frames \cite{Fuentes-Schuller:2004iaz,Alsing2006}  helped popularize RQI in the 21st century. Since then, tools from quantum information have continued to be used to help explain the interface between relativistic and quantum physics, broadening the scope of RQI to include entanglement harvesting protocols \cite{Salton:2014jaa,Pozas-Kerstjens:2015gta,Teixido-Bonfill:2025wqb}, gravitationally-induced entanglement \cite{Marletto:2017kzi,Bose:2017nin}, and, most recently, QRFs \cite{barbado_unruh_2020}.

Here, we consider Alice and Rob to be non-inertial observers of a fermionic field and find that the perspectival entanglement $\EE_p$ degrades much like in the standard global set-up \cite{Alsing2006}. However, this degradation of entanglement $\EE_p$ between Alice and Rob can be completely offset by accounting for the coherence $\CC_p$ in the given perspective, so that the sum $\EE_p+\CC_p$ is constant for all accelerations. This result can then be related to entanglement transference by noting that the standard quantum state used in entanglement degradation \cite{Alsing2006} is a global even state.

From here, one is naturally interested in the interface between QRFs and quantum gravity. This interface has been studied in a number of contexts \cite{Giacomini:2021aof,Kabel:2022cje,delaHamette:2021iwx,Castro-Ruiz:2019nnl,Apadula:2022pxk}. Recently, it was shown that the sum of entanglement and subsystem coherence can be understood as an invariant quantity under perspectival QRF transformations \cite{Cepollaro2024}. It is not yet apparent how this result, when applied to qubits, could be modified by acceleration or spacetime curvature. In this paper, we confirm that the result in \cite{Cepollaro2024} holds for the non-inertial case of entanglement degradation. The non-inertial problem may provide clues for a full generalization of \cite{Cepollaro2024} and, more broadly, of perspectival QRFs, to curved spacetime.

Our paper is organized as follows. In Section \ref{sec:formalism}, we present the formalism that is used to understand our results, including perspectival QRFs, the assignment procedure, and the setup for the entanglement degradation problem in a fermionic field. In Section \ref{sec:transference}, we discuss entanglement transference and parity states. In Section \ref{sec:qrf-deg}, we revisit entanglement degradation in the perspectival picture. Finally, in Section \ref{sec:discussion}, we comment on our findings and draw conclusions. Appendices \ref{sec:app-qrf} through to \ref{sec:app-lin-entropy} supplement the discussion in Sections \ref{sec:formalism}, \ref{sec:transference}, and \ref{sec:qrf-deg}, while Appendix \ref{sec:app-mutual-info} provides a correction to the mutual information in the standard entanglement degradation setting \cite{Alsing2006} and provides the perspectival extension.

\section{FORMALISM} \label{sec:formalism}

In this section, we present the mathematical formalism required for our work. First, we present the notion of perspectival quantum reference frames, including a new methodology for assigning a quantum reference frame (i.e., assigning a perspective) from a global state. Then we introduce some ideas about quantum fields in curved spacetime, required in order for us to apply our quantum reference frame results to the problem of entanglement degradation.

\subsection{Perspectival quantum reference frames} \label{ssec:qrfs}

Perspectival quantum reference frames  \cite{Giacomini:2017zju,delaHamette:2020dyi} are a tool to describe a multipartite quantum system 
$\ket{\psi} \in \mathcal{H} = \mathcal{H}_A\otimes\mathcal{H}_B\otimes \dots$
relative to its subsystems ($A$, $B$, \dots). 
A \emph{quantum reference frame transformation} can then be understood as a transformation from a first description of the multipartite system from the perspective of a given subsystem $A$ to a different description of the same multipartite system from the perspective of a different subsystem $B$.
We will use the superscript bracket notation to describe the subsystem with respect to whose perspective the joint state is being described. For example, $\ket{\psi}^{(A)}$ describes the state $\ket{\psi}$ with respect to the perspective of $A$.

In its general form, the perspectival approach to QRFs can be defined for a group $G$ with the associated Hilbert space $\mathcal{H}=L^2(G)$ \cite{delaHamette:2020dyi}.
This Hilbert space can then be spanned by an orthonormal basis $\{\ket{g}\}$ defined by the group elements $g\in G$.

The most general QRF transformation from a perspective $B$ to a perspective $A$ can then be written as
\begin{equation}
\label{eq:qrf-transformation}
    S^{(B)\rightarrow(A)} = \int_G dg \ket{g^{-1}}_{B\, A}\negthinspace\bra{g}\otimes U^\dagger(g),
\end{equation}
where $U(g)$, acting on the subsystems beyond $A$ and $B$, is the left regular representation of $G$, such that $U(g)\ket{g'}=\ket{g\circ g'}$.
(For a discrete group, the integral is replaced by a sum, and the Hilbert space $L^2(G)$ by $\mathbb{C}[G]$).

\subsubsection{Perspective assignment}
\label{sssec:qrf-persepctives}

In the existing perspectival QRF literature, there is an implicit assumption that in order to talk about QRF transformations, one must be provided with a \emph{perspectival state}, which is already describing the global quantum system from the perspective of one of its sub-systems.
There is a tacit understanding of how one might reinterpret such a global state as a perspectival state, but to the best of our knowledge, this procedure has not been made mathematically rigorous.

Here, we provide the standard intuitive explanation of this procedure in the case of an $N$-qubit pure state,
where we use the symmetry group $\mathbb{Z}_2$ to describe our qubits (which can be thought of as spin-1/2 systems).
We reframe this perspective-assigning procedure in terms of the standard quantum information language of dephasing and purification operators in Appendix \ref{sec:app-qrf}.

Given an $N$-qubit pure state, we define the perspective assigning operation $\mathcal{R}_{P}:(\mathbb{C}^2)^{\otimes N}\rightarrow (\mathbb{C}^2)^{\otimes (N-1)}$, taking an initial $N$-qubit state to the $(N-1)$-qubit state from the perspective of one of the qubits $P$. 
First, we must express the global $N$-qubit state in some physically relevant basis. 
Then, provided with the state representation in the preferred basis, and given the subsystem $P$ whose perspective we want to consider, we perform the following transformations.

For each basis state with the target qubit $P$ in the state $\ket{0}_P$, we leave the basis state unchanged. For each basis state with the target qubit $P$ in the state $\ket{1}_P$, we flip all the qubits (resulting in a basis state with $\ket{0}_P$). 
In general, this procedure maps $2^N$ superposed basis state terms into $2^{N-1}$ pairs of repeated basis states. In order to recover a valid quantum state, we must then renormalize the coefficients in front each basis state.

Given a pair of basis states related by a flip of each qubit with coefficients $c_1$ and $c_2$ (the flip of each qubit means that the two basis states will map to the same repeated pair), one may define the new coefficients as
\begin{align} \label{eq:combination}
    \sqrt{|c_1|^2+|c_2|^2}.
\end{align}

At this point, the state $\ket{0}_P$ factorizes out, and we omit it altogether, thus obtaining a state of $(N-1)$ subsystems from the perspective of $P$.

\subsubsection{Example}
\label{sssec:qrf-example}

As an example, let us consider the initial state of three qubits $A$, $B$, and $C$ expressed as
\begin{align}
    \ket{\psi}_{ABC}=\frac{1}{2}\rr{\ket{000}+\ket{001}+\ket{010}+\ket{111}}.
\end{align}

If we want to obtain the state in the perspective of $P=B$, we must arrange for subsystem $B$ to be in the state $\ket{0}_B$ and separable from the rest. In order to achieve this, we apply the protocol described above.

Flipping all of the qubits for every basis state with $\ket{1}_B$ (namely the third and fourth basis states above) results in the unnormalized vector
\begin{align}
    \ket{\tilde\psi}^{(B)}_{ABC}=\frac{1}{2}\rr{\ket{000}+\ket{001}+\ket{101}+\ket{000}}.
\end{align}

Since the $\ket{000}$ term is repeated, we renormalize the associated coefficient to be $\sqrt{|1/2|^2+|1/2|^2}=\frac{1}{\sqrt{2}}$. This leaves us with the normalized state
\begin{align}
    \ket{\psi}^{(B)}_{ABC}=\frac{1}{\sqrt{2}}\ket{000}+\frac{1}{2}\ket{001}+\frac{1}{2}\ket{101}.
\end{align}

Factoring out $\ket{0}_B$, we obtain the perspectival state
\begin{align}
    \ket{\psi}^{(B)}_{AC}=\mathcal{R}_B\ket{\psi}_{ABC}=\frac{1}{\sqrt{2}}\ket{00}+\frac{1}{2}\ket{01}+\frac{1}{2}\ket{11}.
\end{align}

There are a few important observations concerning the perspective assignment operation $\mathcal{R}_P$. First, $\mathcal{R}_P$ is not unitary. This becomes clear by noting that the reduction in the degrees of freedom of the input state is analogous to a partial trace. Second, we rely on the assumption that a qubit always sees itself in the $\ket{0}$ state, and never in superposition. This convention for perspectival QRFs goes beyond the postulates of quantum mechanics.
Third, we employ the group structure $G=\mathbb{Z}_2$ to describe our qubits via the map $0\mapsto\ket{0},~1\mapsto\ket{1}$. While this is in keeping with the perspectival quantum reference frame literature \cite{delaHamette:2020dyi,Cepollaro2024}, it results in the absence of a complex phase in the perspectival state coefficients.
One might anticipate that the more general form of Eq. \eqref{eq:combination} would be $\sqrt{|c_1|^2+|c_2|^2}\exp\rr{i\varphi(c_1,c_2)}$. 
The simplest assumption, which we take, is that $\varphi(c_1,c_2)=0$.
Conservatively, we can say that the convention used to define $\mathcal{R}_P$ is fully intuitive in the rebit case (i.e., $\mathcal{R}_P:(\mathbb{R}^2)^{\otimes N}\rightarrow(\mathbb{R}^2)^{\otimes N-1}$). However, we find that our results extend to complex inputs, \emph{even when using the real-valued $\mathcal{R}_P$ to obtain perspectival states}.

\subsubsection{Perspectival quantum resources}
\label{sssec:qrf-E-plus-C}

It is well-known that certain quantum resources, such as entanglement and coherence, are not preserved by QRF transformations, as first noted in \cite{Giacomini:2017zju}. For a sufficiently simple system, however, the sum of entanglement and subsystem coherence is invariant under quantum reference frame transformations \cite{Cepollaro2024}. (A \emph{sufficiently simple} system is a pure state comprised of three subsystems with identical Hilbert spaces admitting a group structure as described in the Section \ref{ssec:qrfs} introduction.)

In particular, if we quantify the entanglement and subsystem coherence using the measures of entanglement entropy
defined as $\EE (\ket{\psi}_{AB})=\mathcal{S}(\rho_A)$, where $\rho_A=\Tr_B(\ket{\psi}_{AB\, AB}\negthinspace\bra{\psi})$ is the reduced density matrix of system $A$ and $\mathcal{S}(\rho)=-\Tr(\rho\log\rho)$ is the von Neumann entropy,
and relative entropy of subsystem coherence defined as $\CC(\rho)=\mathcal{S}(\rho)-\mathcal{S}(\rho_d)$, where $\rho_d$ is the state consisting of only the diagonal elements of $\rho$ in the group element basis $\{\ket{g}\}_{g\in G}$.
Then their sum $\EE+\CC$, when appropriately defined on the subsystems, is invariant under QRF transformations\footnote{We will show later  that for each perspective there is a choice of two possible subsystem coherences. In order to recover the invariant quantity, we must choose the correct one of these.}.

Consider, from the perspective of $B$, the entanglement  $\EE(\ket{\psi}_{AC}^{(B)})\equiv \EE_{A,C}^{(B)}$ and the coherence of subsystem $A$, $\CC(\rho_A^{(B)})\equiv \CC_{A}^{(B)}$, where $\rho_A^{(B)}=\Tr_C(\rho_{AC}^{(B)})$, and define things analogously from the perspective of $A$.
It has been shown \cite{Cepollaro2024} that the sum of entanglement and subsystem coherence is invariant, for an appropriately simple system, as expressed by
\begin{equation}\label{eq:e-c-gen}
    \EE_{A,C}^{(B)} + \CC_{A}^{(B)} = \EE_{B,C}^{(A)} + \CC_{B}^{(A)}.
\end{equation}

Alternatively, we may define $\EE^{(B)}_{A,C}\equiv\EE_\ell(|\psi\rangle^{(B)}_{AC})=1-\mathrm{Tr}({\rho^{(B)~2}_{A}})$ (i.e., the linear entropy) and $\CC^{(B)}_A\equiv\CC_{\ell^2}(\rho^{(B)}_A)=\sum_{i\neq j}|\rho^{(B)}_{A,ij}|^2$ (i.e., the $\ell^2$-norm of coherence), and Eq. \eqref{eq:e-c-gen} would hold for these  quantifiers as well.

\subsection{Quantum fermionic state} \label{ssec:quantum-state}

\begin{figure}
    \centering
    \includegraphics[width=1\linewidth]{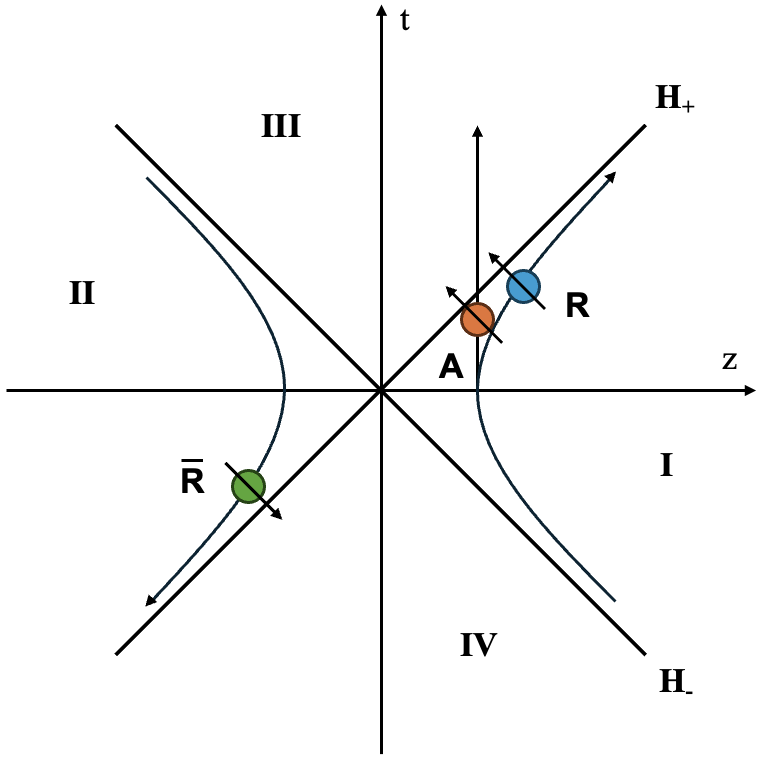}
    \caption{Schematic diagram of Alice, Rob, and anti-Rob's qubits. Alice (A) is inertial, while Rob (R) accelerates uniformly in region I. Another observer, anti-Rob ($\bar R$), travels along a complementary hyperbola in region II. Alice and Rob's particle detectors, which measure a Dirac field, are assumed to be sensitive to the same frequency.}
    \label{fig:rindler-diagram}
\end{figure}

To study the effect of non-inertial frames on QRFs, we consider the setup first proposed by Alsing et al. in \cite{Alsing2006} (see also Appendix \ref{sec:app-dirac}). A brief summary is provided here and in Fig. \ref{fig:rindler-diagram}.

We begin with a free  Dirac field in (3+1)-dimensional Minkowski spacetime  satisfying the field equation,
\begin{align}
    \rr{i\gamma^\mu\partial_\mu-m}\psi=0,
\end{align}
with particle mass $m$ and Dirac matrices $\gamma^\mu$, setting $c=\hbar=1$. The spinor wavefunction $\psi$ may be expanded over positive (fermionic) and negative (antifermionic) modes, $\psi_k^+$ and $\psi_k^-$ respectively. That is,
\begin{equation}
    \psi=\int dk\rr{a_k\psi_k^+ +b_k^\dagger\psi_k^-},
\end{equation}
where $\rr{\psi_k^\pm}_s=(2\pi\omega_k)^{-1/2}\phi_s^\pm e^{\pm i(\mathbf{k\cdot x}-\omega_k t)}$, $\omega_k=(m^2+\mathbf{k}^2)^{1/2}$, and $\phi_s=\phi_s(\mathbf{k})$ is a constant spinor with $s=\{\uparrow,\downarrow\}$ indicating spin-up or spin-down along the quantization axis. Since $\psi$ is a Dirac field, $(a_k^\dagger)^2=(b_k^\dagger)^2=0$. We write down a maximally entangled state of two fermionic modes,
\begin{equation} \label{eq:initial-state}
    \ket{\phi_{k_A,k_R}} = \frac{1}{\sqrt{2}} (\ket{0_{k_A}}^+\ket{0_{k_R}}^++\ket{1_{k_A}}^+\ket{1_{k_R}}^+).
\end{equation}

We adopt the convention in \cite{Alsing2006}  that the first mode is observed by Alice and the second mode by Rob; Alice and Rob have monochromatic detectors sensitive to modes $k_A$ and $k_R$ respectively. In Rindler spacetime, we may expand $\psi$ in the right and left modes, denoted $\psi_k^{R\pm}$ and $\psi_k^{\bar R\pm}$;
\begin{equation}
    \psi=\int dk\rr{c_k^R\psi_k^{R+}+d_k^{R\dagger}\psi_k^{R-}+c_k^{\bar R}\psi_k^{\bar R+}+d_k^{\bar R\dagger}\psi_k^{\bar R-}},
\end{equation}
where $c_k^{R,\bar R}$ and $d_k^{R,\bar R}$ satisfy the appropriate Bogoliubov transformations~\cite{Alsing2006} in relation to $a_k$ and $b_k$. 
We assume that Rob is uniformly accelerating and thus observes Rindler modes. Specifically, we take the ansatz that
\begin{align}
    |0_{k_R}\rangle^+\simeq|0_{k}\rangle^+=\sum_{n=0}^1 A_n|n_{k}\rangle_R^+|n_{-k}\rangle_{\bar R}^-,
\end{align}
with
\begin{align}
\begin{tabular}{l l}
    $c_k^R|0_k\rangle_R^+=0$ & \quad $d_{-k}^{\bar R}|0_{-k}\rangle_{\bar R}^-=0,$ \\
    $c_k^{R\dagger}|0_k\rangle_R^+=|1_k\rangle_R^+$ & \quad $d_{-k}^{\bar R\dagger}|0_{-k}\rangle_{\bar R}^-=|1_{-k}\rangle_{\bar R}^-,$
\end{tabular}
\end{align}
where the subscripts $R$ and $\bar R$, again, differentiate Fock states in the right and left Rindler wedges respectively.
We have assumed at this point that $k_R\approx k_A=k$ while also going from Minkowski to Rindler, which is loosely referred to in the literature as the single-mode approximation\footnote{To be precise, the single mode approximation is unphysical, since a single Minkowski mode does not transform into a single Rindler mode, but rather a highly non-monochromatic field excitation \cite{Eduardo:2010}. What we mean is that we begin with a Minkowski wavepacket that transforms into a suitably narrow Rindler wavepacket centered about the wavenumber $k$, which we then approximate to be monochromatic.}. 
With some computation, we find that
\begin{align}
    &|0_k\rangle^+=\cos r|0_k\rangle_R^+|0_{-k}\rangle_{\bar R}^- + \sin r|1_k\rangle_R^+|1_{-k}\rangle_{\bar R}^-, \label{eq:zeroM} \\
    &|1_k\rangle^+=|1_k\rangle_R^+|0_{-k}\rangle_{\bar R}^-, \label{eq:oneM}
\end{align}
where $\tan r=\exp(-\pi\omega/a)$ and $r\in(0,\pi/4)$ grows monotonically with $a\in (0,\infty)$. Using Eq. \eqref{eq:zeroM}-\eqref{eq:oneM}, we then rewrite Eq. \eqref{eq:initial-state} as
\begin{align} \label{eq:global-state}
    |\phi_r\rangle_{AR\bar R}&=\frac{1}{\sqrt{2}}(\cos r|000\rangle+\sin r|011\rangle+|110\rangle),
\end{align}
where $|000\rangle=|0_k\rangle_A^+|0_k\rangle_R^+|0_{-k}\rangle_{\bar R}^-$, etc. is implied\footnote{In general, fermionic modes do not map to qubits due to the anti-commuting nature of the field, which results in a sign ambiguity \cite{Montero:2011ai,Friis:2013mya}. We enforce the tensor product structure (from left to right) $A_\uparrow$, $A_\downarrow$, $R_\uparrow$, $R_\downarrow$, ${\bar R}_\uparrow$, and then ${\bar R}_\downarrow$ (i.e., order first the systems $A$, $R$, $\bar R$, then, within each system, place the spin-up constituent to the left of the spin-down constituent).}. In Section \ref{sec:qrf-deg}, we will examine the quantum properties of $|\phi_r\rangle_{AR\bar R}$. First, we use our assignment procedure from Section \ref{ssec:qrfs} to obtain
\begin{align}
    &|\phi_r\rangle^{(A)}_{R\bar R}=\frac{1}{\sqrt{2}}(\cos r|00\rangle+\sin r|11\rangle+|01\rangle), \label{eq:persp-A} \\
    &|\phi_r\rangle^{(R)}_{A\bar R}=\frac{1}{\sqrt{2}}(\cos r|00\rangle+\sin r|10\rangle+|01\rangle), \label{eq:persp-R} \\
    &|\phi_r\rangle^{(\bar R)}_{AR}=\frac{1}{\sqrt{2}}(\cos r|00\rangle+\sin r|10\rangle+|11\rangle), \label{eq:persp-Rbar}
\end{align}
which are the two-qubit states as seen from the perspective of the observer in the superscript. We note that these states can be related to each other by the standard quantum reference frame transformations given in Eq. \eqref{eq:qrf-transformation} for $G=SU(2)$.

\section{ENTANGLEMENT TRANSFERENCE} \label{sec:transference}

The statement of entanglement transference is as follows: for a global, 3-qubit pure state $\ket{\psi}_{ABC}$, entanglement transference holds if
\begin{align} \label{eq:conjecture}
    \EE^{(\alpha)}_{\beta,\gamma}+\CC^{(\alpha)}_\beta=\EE_{\gamma,\alpha\beta}
\end{align}
is satisfied simultaneously for $\alpha$, $\beta$, $\gamma$ over all permutations of $A$, $B$, $C$, where $\EE$ and $\CC$ are measures of entanglement and coherence. Eq. \eqref{eq:conjecture} is six equations, although we will show that only three are independent; one of the equations, for example, is
\begin{align} \label{eq:conjecture-example}
    \EE^{(A)}_{B,C}+\CC^{(A)}_B=\EE_{C,AB}.
\end{align}

The right hand side of Eq. \eqref{eq:conjecture-example}, $\EE_{C,AB}$, does not have a superscript, because it is a global quantity. Specifically, it is the bipartite entanglement between one system, in this case $C$, with the other two, $A$ and $B$, treated jointly. Eq. \eqref{eq:conjecture-example} states that entanglement plus subsystem coherence in one perspective is equal to non-perspectival entanglement for a corresponding bipartition of the three systems. The arrangement of the labels $A$, $B$, and $C$ can be remembered using the following aid: repeated labels on the perspectival side of the equation form the joint system within the bipartition, while the label that only appears once on the perspectival side stands alone in $\EE_{\gamma,\alpha\beta}$.

Since $\EE_{\gamma,\alpha\beta}=\EE_{\gamma,\beta\alpha}$, we obtain three indenpendent equations for entanglement transference and not six, as one might naively assume. Eq. \eqref{eq:conjecture} thus contains as a corollary
\begin{align} \label{eq:conjecture-corollary}
    \EE^{(\alpha)}_{\beta,\gamma}+\CC^{(\alpha)}_\beta=\EE^{(\beta)}_{\alpha,\gamma}+\CC^{(\beta)}_\alpha,
\end{align}
which is exactly the set of conditions that was investigated in \cite{Cepollaro2024}. However, we note that Eq. \eqref{eq:conjecture} describing entanglement transference is a stronger condition than Eq. \eqref{eq:conjecture-corollary}, since one can satisfy the latter while violating the former. An example is the global state
\begin{multline} \label{eq:counterexample}
    \ket{\psi}_{ABC}=\ket{0}_A\otimes\frac{1}{\sqrt{2}}\rr{\ket{0}_B+\ket{1}_B} \\
    \otimes\frac{1}{\sqrt{2}}\rr{\ket{0}_C+\ket{1}_C}.
\end{multline}
Clearly, $\EE_{C,AB}=0$, as this state is fully separable. Since $A$ is already in the $\ket{0}_A$ state, we may read $\ket{\psi}^{(A)}_{BC}$ directly from \eqref{eq:counterexample} by ignoring the first system. $\ket{\psi}^{(A)}_{BC}$ is also separable, and so $\EE^{(A)}_{BC}=0$. However, both subsystems from $A$'s perspective have maximal coherence. Hence, $\EE^{(A)}_{B,C}+\CC^{(A)}_B\neq\EE_{C,AB}$. At the same time, it is possible to fully satisfy Eq. \eqref{eq:conjecture-corollary} for this state. This follows from using the assignment procedure in Section \ref{ssec:qrfs} and then applying the result in \cite{Cepollaro2024}.

Identifying the states for which Eq. \eqref{eq:conjecture} holds is, in general, difficult, in part because solutions depend on the measures of $\EE$ and $\CC$ that are chosen and the assignment procedure that maps a global state to its perspectival states. However, using the assignment procedure defined in Section \ref{ssec:qrfs}, with $\EE$ and $\CC$ measured by either (1) entanglement entropy and entropy of coherence or (2) linear entropy and $\ell^2$-norm of coherence, there exists a natural class of sufficient solutions, which we call \emph{parity} states. This description encompasses two sets of simultaneous solutions, the \emph{even} and \emph{odd} states, given by
\begin{align}
    &\ket{E}=E_1\ket{000}+E_2\ket{011}+E_3\ket{101}+
    E_4\ket{110}, \label{eq:even} \\
    &\ket{O}=O_1\ket{001}+O_2\ket{010}+O_3\ket{100}+
    O_4\ket{111}. \label{eq:odd}
\end{align}

\begin{figure}[h]
    \centering
    \includegraphics[width=1\linewidth]{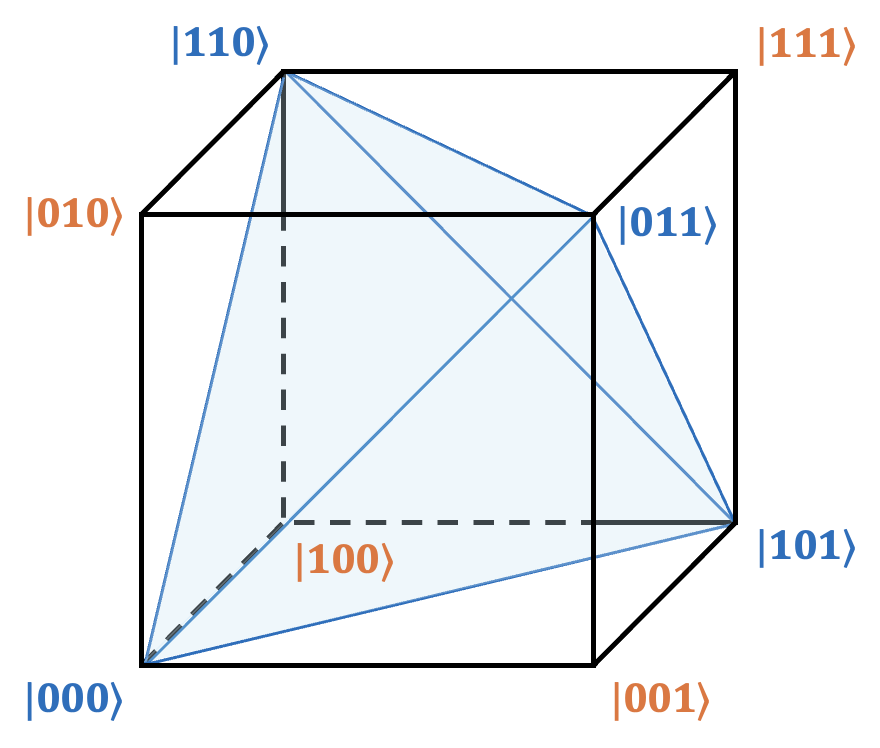}
    \caption{Visual representation of the even and odd parity states, $\ket{E}$ (blue) and $\ket{O}$ (orange) respectively. The tetrahedral arrangement is drawn in explicitly for $\ket{E}$. All vertices for a given parity state are separated by Hamming distance 2.}
    \label{fig:parity-sets}
\end{figure}

\newpage
That is, $\ket{E}$ and $\ket{O}$ satisfy all three conditions of Eq. \eqref{eq:conjecture} simultaneously. The parity states are maximally unconstrained solutions in that one can select any complex coefficients $E_i$ or $O_i$, as long as the state is normalized. Hence, the dimensionality of each solution space is $2\cdot2^2-1=7$. In Appendix \ref{sec:app-math-details}, we show that $\ket{E}$ and $\ket{O}$ satisfy the entanglement transference constraints for entanglement entropy, and we discuss other mathematical details of the constraint problem. The linear entropy case is covered in Appendix \ref{sec:app-lin-entropy}.

The solutions $\ket{E}$ and $\ket{O}$
can be geometrically visualized, which we
depict in Fig. \ref{fig:parity-sets}.
Each parity state forms a tetrahedron, where the vertices are the basis states with non-vanishing coefficients. Notably, the vertices are always separated by a Hamming distance of 2. Well-known states that satisfy Eq. \eqref{eq:conjecture} are the generalized W states $\ket{W_\mathrm{even}}=w_1\ket{011}+w_2\ket{101}+w_3\ket{110}$ and $\ket{W_\mathrm{odd}}=z_1\ket{001}+z_2\ket{010}+z_3\ket{100}$, which are subsets of the parity states respectively. 
The GHZ class of states, $\ket{GHZ}=g\ket{000}+\sqrt{1-g^2}e^{i\varphi}\ket{111}$, $g\in (0,1)$, $\varphi\in[0,2\pi)$, can be shown to violate all cases of Eq. \eqref{eq:conjecture}.

Let us conclude this section with a brief comment on the term \emph{entanglement transference}. If we rearrange Eq. \eqref{eq:conjecture} to read
\begin{align} \label{eq:conjecture-rearranged}
    \CC^{(\alpha)}_\beta=\EE_{\gamma,\alpha\beta}-\EE^{(\alpha)}_{\beta,\gamma},
\end{align}
then Eq. \eqref{eq:conjecture-rearranged} tells us that (1) $\EE^{(\alpha)}_{\beta,\gamma}$, the perspectival entanglement, is always less than $\EE_{\gamma,\alpha\beta}$, and that (2) the difference between the two entanglement quantities is the subsystem coherence. Thus, part of the global entanglement transfers to perspectival coherence, while the rest remains as perspectival entanglement.

\section{ENTANGLEMENT DEGRADATION AND QRFS} \label{sec:qrf-deg}

\begin{figure}
    \centering
     \includegraphics[width=1\linewidth]{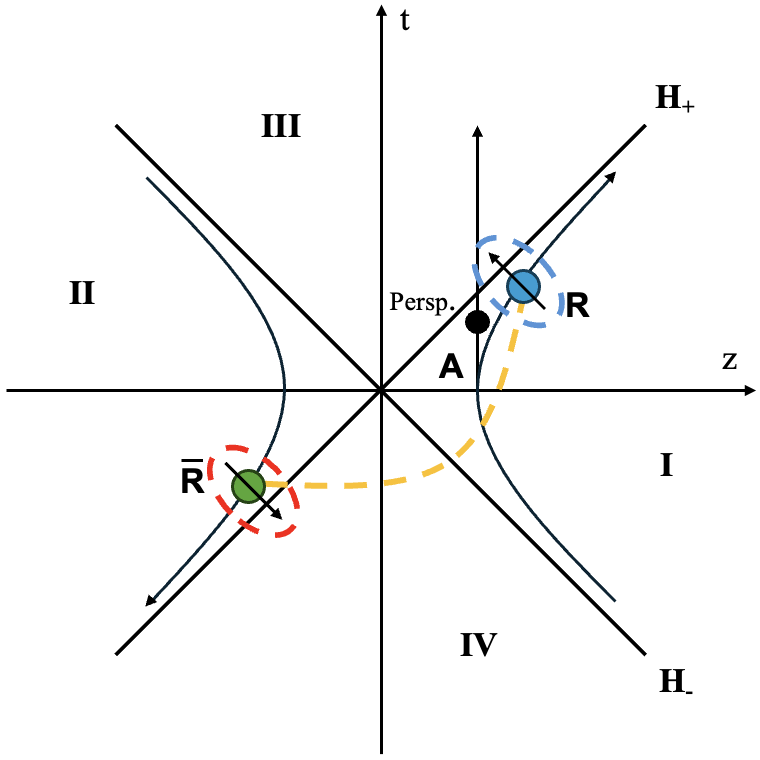}
    \includegraphics[width=1\linewidth]{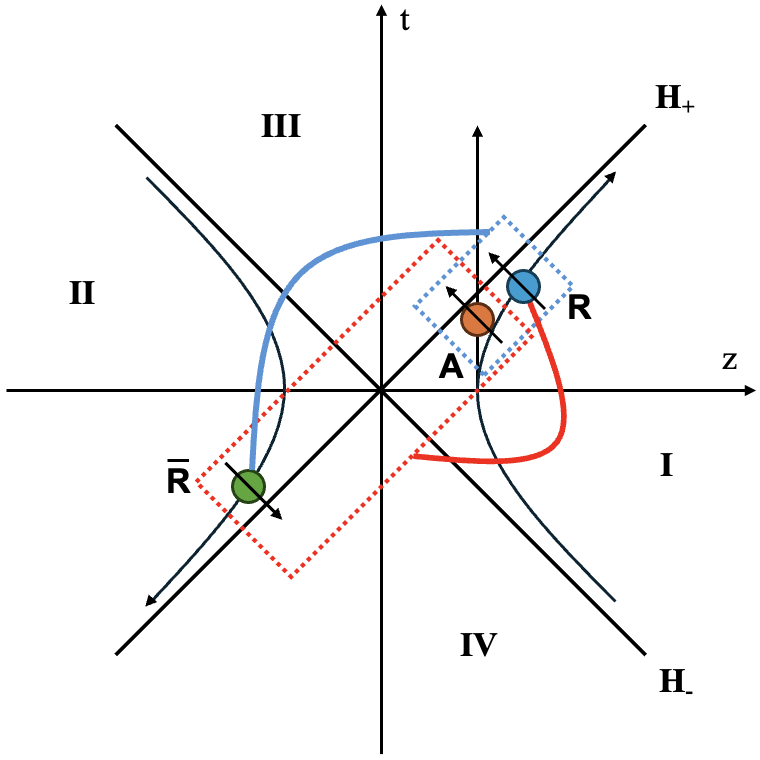}
    \caption{A pictorial representation of entanglement transference relating the perspectival entanglement and coherence (top) to the global entanglement (bottom) for the case of $\EE^{(A)}_{R,\bar R}+\CC^{(A)}_{R}=\EE_{\bar R,AR}$ and $\EE^{(A)}_{R,\bar R}+\CC^{(A)}_{\bar R}=\EE_{R,A\bar R}$. The top diagram shows the perspective of Alice's qubit, with the entanglement $\EE^{(A)}_{R,\bar R}$ depicted by the yellow dashed line connecting Rob and anti-Rob's qubits. The coherences $\CC^{(A)}_{R}$ and $\CC^{(A)}_{\bar R}$ are represented by blue and red dashed ovals around the Rob's and anti-Rob's qubits respectively. The bottom diagram shows two distinct bipartitions of the 3-qubit state $\ket{\psi}_{AR\bar R}$ for which entanglement is calculated. The entanglement between $\bar R$ and the joint system $AR$, $\EE_{\bar R,AR}$, is shown in blue, while the entanglement between $R$ and the joint system $A\bar R$, $\EE_{R,A\bar R}$, is shown in red. 
    The yellow dashed quantity plus the blue (red) dashed quantity in the top diagram equals the blue (red) solid quantity in the bottom one.}
    \label{fig:2e-2c-equals-3e}
\end{figure}

In this section, we revisit the problem of entanglement degradation for a fermionic field \cite{Alsing2006} from the perspective of QRFs.
As before, we use the assignment procedure from Section \ref{ssec:qrfs}. We consider the case where $(\EE,\CC)$ are entanglement entropy and entropy of coherence respectively. The analogous results for linear entropy and $\ell^2$-norm of coherence are presented in Appendix \ref{sec:app-lin-entropy}. 

\begin{table*}
    \centering
    \begin{tabular}{|c|c|c|}
        \hline
        Quantity & Expression & Scope \\
        \hline
        $\EE^{(A)}_{R,\bar R}$ & $-\frac{1}{2}\sum_{j=1}^2\rr{1+(-1)^j\frac{\sqrt{7+\cos(4r)}}{2\sqrt{2}}}\log_2\left[\frac{1}{2}\rr{1+(-1)^j\frac{\sqrt{7+\cos(4r)}}{2\sqrt{2}}}\right]$ &  \\
        $\EE^{(R)}_{A,\bar R}$ & $-\sum_{j=1}^2\rr{\frac{1+(-1)^j\cos r}{2}}\log_2\left[\frac{1+(-1)^j\cos r}{2}\right]$ & Perspectival \\
        $\EE^{(\bar R)}_{A,R}$ & $-\sum_{j=1}^2\rr{\frac{1+(-1)^j\sin r}{2}}\log_2\left[\frac{1+(-1)^j\sin r}{2}\right]$ & \\
        \hline 
        $\EE_{\bar R,AR}$ & $-\sum_{j=1}^2\rr{\frac{1+(-1)^j\cos^2r}{2}}\log_2\left[\frac{1+(-1)^j\cos^2r}{2}\right]$ & \\
        $\EE_{R,A\bar R}$ & $-\frac{\cos^2 r}{2}\log_2\left[\frac{\cos^2 r}{2}\right]-\rr{1-\frac{\cos^2 r}{2}}\log_2\left[1-\frac{\cos^2 r}{2}\right]$ & Global \\
        $\EE_{A,R\bar R}$ & 1 &  \\
        \hline
    \end{tabular}
    \caption{Entanglement characterization of the state $\ket{\phi_r}_{AR\bar R}$, where $\EE$ is entanglement entropy.}
    \label{tab:entanglement1}
\end{table*}

The fermionic state \eqref{eq:global-state} used to study entanglement degradation,
\begin{align}
    |\phi_r\rangle_{AR\bar R}&=\frac{1}{\sqrt{2}}(\cos r|000\rangle+\sin r|011\rangle+|110\rangle), \nonumber
\end{align}
is a global even state. Thus, it satisfies the entanglement transference conditions in Eq. \eqref{eq:conjecture}. Two of these conditions are illustrated in Fig. \ref{fig:2e-2c-equals-3e}. Since $\ket{\phi_r}_{AR\bar R}$ satisfies Eq. \eqref{eq:conjecture}, the $\EE+\CC$ properties of the system can be fully characterized by providing six entanglement quantities, which we list in Table \ref{tab:entanglement1}.

The invariant quantity $\EE+\CC$ identified in \cite{Cepollaro2024}, which is a special case of Eq. \eqref{eq:conjecture-corollary}, gives rise to three equations:
\begin{align}
    \EE^{(A)}_{R,\bar{R}} + \CC^{(A)}_{R} &= \EE^{(R)}_{A,\bar{R}} + \CC^{(R)}_{A}, \label{eq:pair1} \\
    \EE^{(R)}_{A,\bar{R}} + \CC^{(R)}_{\bar{R}} &= \EE^{(\bar{R})}_{A,R} + \CC^{(\bar{R})}_{R}, \label{eq:pair2} \\
    \EE^{(\bar{R})}_{A,R} + \CC^{(\bar{R})}_{A} &= \EE^{(A)}_{R,\bar{R}} + \CC^{(A)}_{\bar{R}}. \label{eq:pair3}
\end{align}

\begin{figure*}
    \centering
    \includegraphics[width=0.32\linewidth]{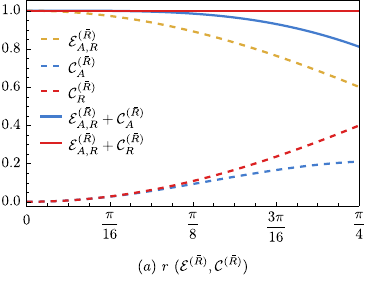}
    \includegraphics[width=0.32\linewidth]{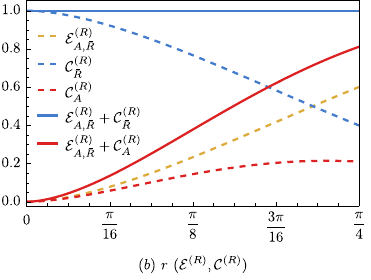}
    \includegraphics[width=0.32\linewidth]{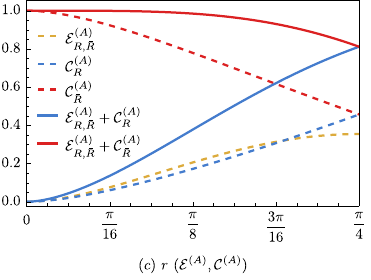}
    \caption{Comparison of entanglement entropy $\EE$ plus subsystem (entropy of) coherence $\CC$ from the perspectives of (a) anti-Rob, (b) Rob, and (c) Alice in the three-qubit system. Entanglement entropy from each perspective is depicted by the yellow dashed line. Blue colour corresponds to the clockwise cyclic direction, defined to be $A\rightarrow R\rightarrow\bar R\rightarrow A$, while red colour corresponds to the anti-clockwise cyclic direction, $A\rightarrow\bar R\rightarrow R\rightarrow A$. For example, in plot (a), the subsystem coherence of Alice from anti-Rob's perspective is given by the blue dashed line, since Alice is cyclically ``downstream'' of anti-Rob. If anti-Rob makes reference to Rob, who is in the reverse direction, then the corresponding subsystem coherence is the red dashed line. The totals $\EE+\CC$ are given in each plot by the solid lines, which follow the same colour convention. Observe that Eqs. \eqref{eq:pair1}-\eqref{eq:pair3} are satisfied, since red solid lines 
    in a given panel are the same as
    the blue solid lines in the (cyclically) right adjacent panel.}
    \label{fig:e-and-c}
\end{figure*}

Indeed, these three equations are satisfied for the state $|\phi_r\rangle_{AR\bar R}$ for any given $r$. Fig. \ref{fig:e-and-c} shows the entanglement entropy $\EE$ and subsystem entropy of coherence $\CC$ of the $|\phi_r\rangle_{AR\bar R}$ system for (a) Alice and Rob's qubits from the perspective of anti-Rob, (b) Alice and anti-Rob's qubits from the perspective of Rob, and (c) Rob and anti-Rob's qubits from the perspective of Alice, as a function of the acceleration parameter $r$.

In addition to satisfying Eq. \eqref{eq:pair1}-\eqref{eq:pair3}, we observe in Fig. \ref{fig:e-and-c}a that the entanglement between Alice and Rob's qubits from the perspective of anti-Rob begins at the maximal value when $r=0$ and decreases monotonically to a fixed non-zero value as $r$ increases. This behaviour is commensurate with  the entanglement degradation observed in \cite{Alsing2006}, where anti-Rob's system is traced out, and one computes entanglement for the mixed state shared by Alice and Rob. In particular, the entanglement of formation for the reduced mixed state computed in \cite{Alsing2006} is equal to the entanglement entropy of the corresponding perspectival state.

Perhaps most interestingly, we observe that this perspectival degradation in entanglement can be \emph{exactly offset} by accounting for
a complementary perspectival coherence, namely $\CC^{(\bar R)}_R$.
In this sense, the missing entanglement can be interpreted as having been converted into subsystem coherence.
Visually, in Fig. \ref{fig:e-and-c}, the yellow dashed curve plus the red dashed curve equals the red solid curve, which is constant at unity. This result can also be understood via entanglement transference by noting that
$\EE_{A,R\bar R}$ is unity for all accelerations.
If we add the other subsystem's coherence, $\CC^{(\bar R)}_A$, the degradation is offset, but only partially.

As a mathematical detail, we observe that, at infinite acceleration, there is symmetry between Rob and anti-Rob's perspectives. If one flips Fig. \ref{fig:e-and-c}b about a vertical line at $r=\pi/8$ and glues the resulting plot to the right edge of Fig. \ref{fig:e-and-c}a, then the curves in Fig. \ref{fig:e-and-c}a would continue into the reflected version of Fig. \ref{fig:e-and-c}b, with red becoming blue and vice versa. 

\section{DISCUSSION} \label{sec:discussion}

In this paper, we set out to better understand how quantum reference frames (QRFs) can be used as a tool for various problems in quantum physics. Focusing on the perspectival approach to QRFs, we identified a new connection between the quantum resources of related global and perspectival states and applied this to the problem of entanglement degradation from relativistic quantum information (RQI).

We constructed a perspective-assigning operation that allows us to identify the perspectival quantum states given a global quantum state. Using this operation, we then described entanglement transference, which states that the global entanglement of a system can be expressed as a sum of entanglement and coherence within a given perspective. This result provides an interesting new interpretation for the invariant sum of entanglement and coherence under QRF transformations discovered in \cite{Cepollaro2024}.
We showed how entanglement transference is robust for even and odd 3-qubit states, although other exotic solutions have not been ruled out.

Futhermore, we applied perspectival QRFs in the RQI setting to better understanding how QRFs behave in non-inertial frames. We found that entanglement degradation is still manifest for perspectival states. However, perspectival coherence can exactly offset this degradation, offering an important insight into the much studied problem. Moreover, we showed that the sum of perspectival entanglement and coherence remains invariant for QRF transformations, even in the presence of non-inertial frames.

The limitations of our study are as follows. First, the results found in this paper were only considered in the context of \emph{perspectival} QRFs with \emph{ideal reference frames}. It would be interesting to investigate how entanglement transference could manifest in other QRF approaches, such as the perspectival-neutral approach for which there exists a concrete connection to the perspectival approach \cite{delaHamette:2021oex}. Likewise, it would be interesting to extend our results to non-ideal QRFs.

Second, the QRF transformations that we considered are decoupled from the motion of the systems,
in that the perspectival formalism does not have an obvious way of accounting for both the state of our qubits \emph{and} the relativistic reference frame that they are in. Further work is needed to more generally understand the effect of relativistic causal structures on perspectival states.

Third, the mapping of fermionic field modes to qubits, although widely used in the RQI literature, is imperfect. We assumed a specific tensor product structure to compensate. However, we note that, even if one flips the signs of any combination of coefficients in the global state, the overall state will still be of even parity, and hence entanglement transference will hold. As discussed in Section \ref{ssec:qrfs}, we are also constrained by a phase ambiguity when defining the perspective-assigning operator due to the $\mathbb{Z}_2$ symmetry group that we consider. 

In the context of RQI, it would be natural to try to extend this study to bosonic modes \cite{Fuentes-Schuller:2004iaz} to see whether the entanglement degradation can also be perfectly offset in that case. We may also consider other tripartite global states, such as those used to investigate entanglement harvesting ~\cite{Lorek:2014dwa,Mendez-Avalos:2022obb}.

It would also be interesting to connect the results in this paper to the broader scope of quantum entanglement theory \cite{Horodecki:2009zz} and of quantum resource theories \cite{Chitambar:2018rnj}. 
How general is the connection between perspectival entanglement entropy and global entanglement of formation observed between this work and \cite{Alsing2006}? Can the phenomena of entanglement transference be related to the SLOCC entanglement classification \cite{Dur:2000zz}?
Can quantum resources besides entanglement and coherence be transferred when considering the perspectival picture? Expanding the connection between quantum information and perspectival QRFs remains an open problem.

\section*{ACKNOWLEDGEMENTS}

 The authors acknowledge constructive discussions with Jack Davis at multiple stages of this project. E. A. P. acknowledges helpful discussions about quantum reference frames with Eduardo Mart\'in-Mart\'inez, Anne-Catherine de la Hamette, and \v{C}aslav Brukner. 
 This work was supported in part by the Natural Sciences and Engineering Research Council of Canada. 
This work was conducted on the traditional territory of the Neutral, Anishnaabeg, and Haudenosaunee Peoples. The University of Waterloo and the Institute for Quantum Computing are situated on the Haldimand Tract, land that was promised to Six Nations, which includes six miles on each side of the Grand River. 


\bibliographystyle{unsrtnat}
\bibliography{ref}

\newpage
\appendix
\begin{widetext}

\section*{APPENDICES}

\section{MORE ON PERSPECTIVAL QUANTUM REFERENCE FRAMES}
\label{sec:app-qrf}


In the existing perspectival quantum reference frame (QRF) literature \cite{Giacomini:2017zju,delaHamette:2020dyi,Cepollaro2024}, there is an implicit assumption that in order to talk about QRF transformations, one must be provided with an initial \emph{perspectival state} that  already describes the global quantum system from the perspective of one of its sub-systems.
There is a tacit understanding of how one might reinterpret such a global state as a perspectival state (as showcased in the example from Section \ref{sssec:qrf-example}). But to the best of our knowledge, this procedure has not been presented mathematically before.

Here, we provide a new mathematical procedure for the assignment of perspective in the case of an $N$-qubit pure state. 
We describe this procedure in the context of standard quantum information theory using notions of quantum channels, dephasing operators and state purification protocols. 
\newline

Given an $N$-qubit pure state, the perspective assigning operation $\mathcal{R}_{P}:(\mathbb{C}^2)^{\otimes N}\rightarrow (\mathbb{C}^2)^{\otimes N-1}$  takes an initial $N$-qubit state to the $(N-1)$-qubit state from the perspective of one of the qubits, $P$.
This operation can be described in 5 steps:
\begin{itemize}
    \item[1.] Express the $N$-qubit pure state as a density matrix, for a preferred basis.
    \item[2.] Apply a generalized dephasing channel to the $N$-qubit system.
    \item[3.] Apply the (non-unitary) \emph{perspective operator} for the $P$-th qubit.
    \item[4.] Trace out the $P$-th qubit, to recover an $(N-1)$-qubit system.
    \item[5.] Purify the resulting perspectival state and project over the ancilla degrees of freedom, to recover the $(N-1)$-qubit system as a pure state.
\end{itemize}

For simplicity and concreteness, we will present the perspective assigning operation for a 3-qubit system. But the following protocol is valid for a general $N$-qubit system. 

Consider the $3$-qubit pure state 
\begin{equation}
    \ket{\psi}_{ABC} = \sum_{i,j,k\in\{0,1\}} a_{ijk} \ket{ijk},
\end{equation}
in a particular basis.

\newpage
(1.) \textbf{Density Matrix.} We can express the state as a density matrix $\rho_{ABC} = \ket{\psi}_{ABC}\leftindex^{}_{ABC}{\bra{\psi}}$.


(2.) \textbf{Dephasing.} We then apply the dephasing operator to our initial state.
The single-qubit dephasing channel has Kraus operators $K_0=\sqrt{1-p}~\bm{1}$ and $K_1=\sqrt{p}~\sigma^z$, where $p$ is the dephasing parameter. Here, we choose $p=1/2$ which corresponds to maximal dephasing.
For 3-qubit dephasing, we define the joint Kraus operators as $K_{ijk}=K_i\otimes K_j\otimes K_k$.

The 3-qubit dephasing operator can then be expressed as $D(\rho_{ABC}) = \sum_{ijk}K_{ijk}\rho_{ABC}K_{ijk}^\dagger$.
The resulting density matrix has the square of the amplitudes $a_{ijk}^2$ along the diagonal and vanishing off-diagonal entries, i.e., the dephased state has no coherence (in the preferred basis).

(3.) \textbf{Perspective Operator} $N^{(P)}$. Once we have the dephased state, we apply the \emph{perspective operator}
\begin{equation}
    N^{(P)} = \ketbra{0}{0}_P\otimes^{N-1}\bm{1}_2 + \ketbra{0}{1}_P\otimes^{N-1}\sigma^x.
\end{equation}

Concretely, if we wanted to implement the perspective of subsystem $A$ to our 3-qubit state, we would have the perspective operator
\begin{equation}
    N^{(A)} = \ketbra{0}{0}_A\otimes\bm{1}_B\otimes\bm{1}_C+\ketbra{0}{1}_A\otimes\sigma^x_B\otimes\sigma^x_C.
\end{equation}

This results in the state
\begin{equation}
    \tilde{\rho}_{ABC}^{(A)} = (a_{000}^2+a_{111}^2)\ketbra{000}{000} + (a_{001}^2+a_{110}^2)\ketbra{001}{001} + (a_{010}^2+a_{101}^2)\ketbra{010}{010} + (a_{011}^2+a_{100}^2)\ketbra{011}{011}.
\end{equation}

(4.) \textbf{Trace out the $p$-th qubit.} We now apply a partial trace on the perspective of the system that has been taken.
Taking the partial trace will leave us with the reduced density matrix with diagonal entries equal to $a_{ijk}^2+a_{(1-i)(1-j)(1-k)}^2$:

\begin{equation}
    \rho_{BC}^{(A)} = (a_{000}^2+a_{111}^2)\ketbra{00}{00} + (a_{001}^2+a_{110}^2)\ketbra{01}{01} + (a_{010}^2+a_{101}^2)\ketbra{10}{10} + (a_{011}^2+a_{100}^2)\ketbra{11}{11}.
\end{equation}

(5.) \textbf{Purification.} Finally, we apply quantum state purification and project over the ancilla degrees of freedom, leaving us with the desired final state. 

Let $\ket{\Psi}^{(A)}_{BC(DE)}$ be the purification given by
\begin{equation}
    \ket{\Psi}^{(A)}_{BC(DE)} = (a_{000}^2+a_{111}^2)^{\frac12}\ket{00}\ket{00} + (a_{001}^2+a_{110}^2)^{\frac12}\ket{01}\ket{01} + (a_{010}^2+a_{101}^2)^{\frac12}\ket{10}\ket{10} + (a_{011}^2+a_{100}^2)^{\frac12}\ket{11}\ket{11},
\end{equation}
with ancilla qubit systems $D$ and $E$.

Project over the ancilla qubit systems to recover the desired perspectival state
\begin{equation}
    \ket{\psi}^{(A)}_{BC} = (a_{000}^2+a_{111}^2)^{\frac12}\ket{00} + (a_{001}^2+a_{110}^2)^{\frac12}\ket{01} + (a_{010}^2+a_{101}^2)^{\frac12}\ket{10} + (a_{011}^2+a_{100}^2)^{\frac12}\ket{11}.
\end{equation}

\section{DIRAC FIELDS AND RINDLER ACCELERATION} \label{sec:app-dirac}

Recall in Section \ref{ssec:quantum-state} the mode expansion
\begin{align}
    \psi=\int dk\rr{a_k\psi_k^+ +a_k^\dagger\psi_k^-},
\end{align}
where $\rr{\psi_k^\pm}_s=(2\pi\omega_k)^{-1/2}\phi_s^\pm e^{\pm i(\mathbf{k\cdot x}-\omega_k t)}$, $\omega_k=(m^2+\mathbf{k}^2)^{1/2}$, and $\phi_s=\phi_s(\mathbf{k})=\{\uparrow,\downarrow\}$. $\phi_s$ satisfies the normalization relations
\begin{align}
    \pm\overline{\phi_s^\pm}\phi_{s'}^\pm=(\omega_k/m)\delta_{ss'}, \quad  \overline{\phi_s^\pm}\phi_{s'}^\pm=0
\end{align}
where the adjoint spinor  $\overline{\phi_s^\pm}=\phi_s^{\pm\dagger}\gamma^0$. Defining the Dirac inner product to be $(\phi(\mathbf{x},t),\varphi(\mathbf{x},t))=\int d\mathbf{x} ~\phi^\dagger(\mathbf{x},t)\varphi(\mathbf{x},t)$,   we obtain the orthonormality conditions
\begin{align}
    (\psi_k^+,\psi_{k'}^+)=-(\psi_k^-,\psi_{k'}^-)=\delta(k-k'), \quad (\psi_k^\pm,\psi_{k'}^\mp)=0.
\end{align}
The future-directed Minkowski Killing vector $\partial_t$ classifies modes as positive frequency ($\psi_k^+\propto e^{-i\omega_k t}$) or negative frequency ($\psi_k^-\propto e^{+i\omega_k t}$), and these correspond to fermions and antifermions respectively. The creation and annihilation operators satisfy the anticommutation relations
\begin{align}
    \{a_i,a_j^\dagger\}=\{b_i,b_j^\dagger\}=\delta_{ij}.
\end{align}

Suppose now that Rob accelerates in the $(t,z)$ Minkowski plane. To cover this plane, we require two sets of Rindler coordinates $(\tau,\zeta)$:
\begin{align}
    &at=e^{a\zeta}\sinh(a\tau), \qquad az=e^{a\zeta}\cosh(a\tau), \quad \mathrm{in~the~right~wedge}, \\
    &at=-e^{a\zeta}\sinh(a\tau), \quad az=-e^{a\zeta}\cosh(a\tau), \quad \mathrm{in~the~left~wedge},
\end{align}
with $a$ denoting Rob's proper acceleration. The basis set of Rindler modes is then $\{\psi_k^{R+},\psi_k^{R-},\psi_k^{\bar R+},\psi_k^{\bar R-}\}$, where the modes labeled by $R$ have support on the right wedge and those labeled by $\bar R$ have support on the left wedge. The notation $\bar R$, implying antiparticle-ness, is justified by the fact that
\begin{align}
    \partial_\tau=\frac{\partial t}{\partial \tau}\partial_t+\frac{\partial z}{\partial \tau}\partial_z=a(z\partial_t+t\partial_z)
\end{align}
is proportional to $\partial_t$ in the right wedge and proportional to $-\partial_t$ in the left wedge. Hence, the future-directed Killing vector in the left wedge is $\partial_{-\tau}=-\partial_\tau$ rather than $\partial_\tau$. As before, $\psi_k^{R+}\propto e^{-i\omega_k \tau}$ is a positive frequency mode in the right wedge, but $\partial_{-\tau}e^{-i\omega_k\tau}=i\omega_k e^{-i\omega_k\tau}$, meaning that $\psi_k^{\bar R-}\propto e^{-i\omega_k\tau}$ is a negative frequency mode in the left wedge. The Rindler modes satisfy the orthonormality relations
\begin{align}
    (\psi_k^{\sigma+},\psi_{k'}^{\sigma'+})=-(\psi_k^{\sigma-},\psi_{k'}^{\sigma'-})=\delta_{\sigma\sigma'}\delta(k-k'), \quad (\psi_k^{\sigma\pm},\psi_{k'}^{\sigma'\mp})=0
\end{align}
for $\sigma,\sigma'\in\{R,\bar R\}$. Recalling from Section \ref{ssec:quantum-state} the expansion
\begin{align}
    \psi=\int dk\rr{c_k^R\psi_k^{R+}+d_k^{R\dagger}\psi_k^{R-}+c_k^{\bar R}\psi_k^{\bar R+}+d_k^{\bar R\dagger}\psi_k^{\bar R-}},
\end{align}
we may compute the Bogoliubov transformations for the single mode approximation to be ~\cite{Takagi:1986,Jauregui:1991,McMahon:2006}
\begin{align}
    &\begin{bmatrix}
        a_k \\ b_{-k}^\dagger
    \end{bmatrix}=
    \begin{bmatrix}
        \cos r & -e^{-i\varphi}\sin r \\
        e^{i\varphi}\sin r & \cos r
    \end{bmatrix}
    \begin{bmatrix}
        c_k^R \\ d_{-k}^{\bar R\dagger}
    \end{bmatrix}, \label{eq:bog1} \\
    &\begin{bmatrix}
        b_k \\ a_{-k}^\dagger
    \end{bmatrix}=
    \begin{bmatrix}
        \cos r & e^{-i\varphi}\sin r \\
        -e^{i\varphi}\sin r & \cos r
    \end{bmatrix}
    \begin{bmatrix}
        d_k^R \\ c_{-k}^{\bar R\dagger}
    \end{bmatrix}, \label{eq:bog2}
\end{align}
where $\tan r=\exp(-\pi\omega/a)$ and $\varphi$ is a phase that can be absorbed into the definition of the creation and annihilation operators. The first set of Bogoliubov transformations \eqref{eq:bog1} mixes a particle in the right wedge with an antiparticle in the left wedge, while the second set \eqref{eq:bog2} mixes an antiparticle in the right wedge with a particle in the left wedge. Note that the Minkowski vacuum for mode $k$ is a two-mode squeezed state in the Rindler description with squeezing operator
\begin{align}
    S=\exp\left[r\rr{c_k^{R\dagger}d_{-k}^{\bar R\dagger}e^{-i\varphi}+c_k^{R}d_{-k}^{\bar R}e^{i\varphi}} \right].
\end{align}
Taking the expression for $a_k$ from the Bogoliubov transformation \eqref{eq:bog1} and substituting the ansatz $|0_{k}\rangle^+=\sum_{n=0}^1 A_n|n_{k}\rangle_R^+|n_{-k}\rangle_{\bar R}^-$ into $a_k|0_k\rangle^+=0$, we obtain Eq. \eqref{eq:zeroM} from Section \ref{ssec:quantum-state}, assuming $\varphi=0$. From there, we apply $a_k^\dagger$ given in \eqref{eq:bog2} to both sides to obtain Eq. \eqref{eq:oneM} from Section \ref{ssec:quantum-state}.

\section{MATHEMATICAL DETAILS ON THE TRANSFERENCE OF ENTANGLEMENT} \label{sec:app-math-details}

Let $\EE$ be entanglement entropy and $\CC$ be entropy of coherence, and consider a general 3-qubit state $\ket{\psi}_{ABC}=a\ket{000}+b\ket{001}+c\ket{010}+d\ket{011}+e\ket{100}+f\ket{101}+g\ket{110}+h\ket{111}$, with the normalization condition $a^2+b^2+c^2+\cdots+h^2=1$. We allow the coefficients $a,b,c,\dots,h$ to be complex, and so $a^2$ should be understood as $|a|^2$. Recall that transference of entanglement can be described by three constraints:
\begin{align}
    &\EE^{(A)}_{B,C}+\CC^{(A)}_B=\EE_{C,AB}, \label{eq:constraint1} \\
    &\EE^{(B)}_{A,C}+\CC^{(B)}_C=\EE_{A,BC}, \label{eq:constraint2} \\
    &\EE^{(C)}_{A,B}+\CC^{(C)}_A=\EE_{B,AC}. \label{eq:constraint3} 
\end{align}

Using the prescription for assigning perspectives described in Section \ref{ssec:qrfs}, the perspectival states are
\begin{align}
    &\ket{\psi}^{(A)}_{BC}=\sqrt{a^2+h^2}\ket{00}+\sqrt{b^2+g^2}\ket{01}+\sqrt{c^2+f^2}\ket{10}+\sqrt{d^2+e^2}\ket{11}, \\
    &\ket{\psi}^{(B)}_{AC}=\sqrt{a^2+h^2}\ket{00}+\sqrt{b^2+g^2}\ket{01}+\sqrt{d^2+e^2}\ket{10}+\sqrt{c^2+f^2}\ket{11}, \\
    &\ket{\psi}^{(C)}_{AB}=\sqrt{a^2+h^2}\ket{00}+\sqrt{c^2+f^2}\ket{01}+\sqrt{d^2+e^2}\ket{10}+\sqrt{b^2+g^2}\ket{11}.
\end{align}

Inserting $\ket{\psi}_{ABC}$ into Eqs. \eqref{eq:constraint1}-\eqref{eq:constraint3}, and recalling that $\CC(\rho)=\mathcal{S}(\rho)-\mathcal{S}(\rho_d)$, we obtain the general form
\begin{align}
    &\EE_{\gamma,\alpha\beta}=-\frac{1-X(\pi)}{2}\log_2\rr{\frac{1-X(\pi)}{2}}-\frac{1+X(\pi)}{2}\log_2\rr{\frac{1+X(\pi)}{2}}, \label{eq:derivation1-3e} \\
    &\EE^{(\alpha)}_{\beta,\gamma}=-\frac{1-Y(\pi)}{2}\log_2\rr{\frac{1-Y(\pi)}{2}}-\frac{1+Y(\pi)}{2}\log_2\rr{\frac{1+Y(\pi)}{2}}, \label{eq:derivation1-2e} \\
    &\CC^{(\alpha)}_\beta=-L(\pi)\log_2(L(\pi))-(1-L(\pi))\log_2(1-L(\pi))-\EE^{(\alpha)}_{\beta,\gamma}, \label{eq:derivation1-2c}
\end{align}
where $\alpha$, $\beta$, $\gamma$ encapsulate all permutations of $A,B,C$ and $X(\pi)$, $Y(\pi)$, and $L(\pi)$ are functions of $a,b,c,\dots,h$ that depend on the permutation, $\pi$. The exact expressions for $X(\pi)$, $Y(\pi)$, and $L(\pi)$ for each permutation are listed in Table \ref{tab:xyl-eece}.

\begin{table}
    \centering
    \begin{tabular}{|c|c|c|}
        \hline
        Constraint & Quantity & Expression \\
        \hline
         & $X(\pi)$ & $\sqrt{1-4(a^2+c^2+e^2+g^2)(b^2+d^2+f^2+h^2)+4|ab^*+cd^*+ef^*+gh^*|^2}$ \\
        \eqref{eq:constraint1} & $Y(\pi)$ & $\sqrt{1-4[(a^2+h^2)(d^2+e^2)+(b^2+g^2)(c^2+f^2)]+8\Pi_p}$ \\
         & $L(\pi)$ & $a^2+b^2+g^2+h^2$ \\
        \hline
         & $X(\pi)$ & $\sqrt{1-4(a^2+b^2+c^2+d^2)(e^2+f^2+g^2+h^2)+4|ae^*+bf^*+cg^*+dh^*|^2}$ \\
        \eqref{eq:constraint2} & $Y(\pi)$ & $\sqrt{1-4[(a^2+h^2)(c^2+f^2)+(b^2+g^2)(d^2+e^2)]+8\Pi_p}$ \\
         & $L(\pi)$ & $a^2+d^2+e^2+h^2$ \\
        \hline
         & $X(\pi)$ & $\sqrt{1-4(a^2+b^2+e^2+f^2)(c^2+d^2+g^2+h^2)+4|ac^*+bd^*+eg^*+fh^*|^2}$ \\
        \eqref{eq:constraint3} & $Y(\pi)$ & $\sqrt{1-4[(a^2+h^2)(b^2+g^2)+(c^2+f^2)(d^2+e^2)]+8\Pi_p}$ \\
         & $L(\pi)$ & $a^2+c^2+f^2+h^2$ \\
        \hline
    \end{tabular}
    \caption{Expressions for $X(\pi)$, $Y(\pi)$, and $L(\pi)$ for entanglement entropy $\EE$ and entropy of coherence $\CC$, where $\Pi_p=\sqrt{(a^2+h^2)(b^2+g^2)(c^2+f^2)(d^2+e^2)}$ is the product of perspectival coefficients.}
    \label{tab:xyl-eece}
\end{table}

Adding Eqs. \eqref{eq:derivation1-2e} and \eqref{eq:derivation1-2c} and comparing the sum to Eq. \eqref{eq:derivation1-3e}, we obtain equality when
\begin{align} \label{eq:cond1}
    L(\pi)=\frac{1-X(\pi)}{2}~\mathrm{ or }~\frac{1+X(\pi)}{2}.
\end{align}

If we take the parity states defined in Eqs. \eqref{eq:even} and \eqref{eq:odd}, we recognize immediately that the cross term in $X(\pi)$ (i.e., the third term under the square root) vanishes for \eqref{eq:constraint1}, \eqref{eq:constraint2}, and \eqref{eq:constraint3}. The remaining terms in $X(\pi)$ form a perfect square. To illustrate the point, consider constraint \eqref{eq:constraint1} for the even state $\ket{\psi}_{ABC}=a\ket{000}+d\ket{011}+f\ket{101}+g\ket{110}$. $L(\pi)=a^2+g^2$ and
\begin{align}
    X(\pi)&=\sqrt{1-4(a^2+g^2)(d^2+f^2)} \\
    &=\sqrt{1-4(a^2+g^2)(1-(a^2+g^2))} \quad\mathrm{by~state~normalization} \\
    &=\sqrt{1-4(a^2+g^2)+4(a^2+g^2)^2} \\
    &=\sqrt{(1-2(a^2+g^2))^2} \\
    &=|1-2L(\pi)|.
\end{align}
Hence, condition \eqref{eq:cond1} is satisfied. For the other constraints, we obtain the same result for any parity state. 

However, it is possible to construct certain (partial) solutions that are not parity states. For example, the state
\begin{align} \label{eq:other-soln}
    \ket{\psi(q)}_{ABC}=\sqrt{\frac{1}{2}-q^2}\ket{000}+q\ket{001}-q\ket{010}+\sqrt{\frac{1}{2}-q^2}\ket{011}, \quad -1/\sqrt{2} < q < 1/\sqrt{2}
\end{align}
satisfies \eqref{eq:constraint1} and \eqref{eq:constraint3}, but not \eqref{eq:constraint2}. Other spaces of solutions (that we have discovered) have lower dimensionality than the space of solutions corresponding to parity states. The solution in Eq. \eqref{eq:other-soln}, for example, has only one degree of freedom, $q\in[-1/2,1/2]$, compared to the full seven degrees of freedom of a parity state---four complex coefficients minus normalization. Additionally, the non-parity solutions do not necessarily solve all three constraints simultaneously. Hence, parity states appear to be the most natural solutions to this set of constraints. 
The full mathematical structure of the space of solutions remains an interesting subject for future study.

\section{LINEAR ENTROPY AND THE \texorpdfstring{$\ell^2$}{ℓ\u00B2}-NORM OF COHERENCE} \label{sec:app-lin-entropy}

\begin{table}[h]
    \centering
    \begin{tabular}{|c|c|c|}
        \hline
        Constraint & Quantity & Expression \\
        \hline
         & $X(\pi)$ & $(a^2+c^2+e^2+g^2)^2+2|ab^*+cd^*+ef^*+gh^*|^2+(b^2+d^2+f^2+h^2)^2$ \\
        \eqref{eq:constraint1} & $Y(\pi)$ & $\big[\sqrt{(a^2+h^2)(c^2+f^2)}+\sqrt{(b^2+g^2)(d^2+e^2)}\big]^2$ \\
         & $L(\pi)$ & $(a^2+b^2+g^2+h^2)^2+(c^2+d^2+e^2+f^2)^2$ \\
        \hline
         & $X(\pi)$ & $(a^2+b^2+c^2+d^2)^2+2|ae^*+bf^*+cg^*+dh^*|^2+(e^2+f^2+g^2+h^2)^2$ \\
        \eqref{eq:constraint2} & $Y(\pi)$ & $\big[\sqrt{(a^2+h^2)(b^2+g^2)}+\sqrt{(c^2+f^2)(d^2+e^2)}\big]^2$ \\
         & $L(\pi)$ & $(a^2+d^2+e^2+h^2)^2+(b^2+c^2+f^2+g^2)^2$ \\
        \hline
         & $X(\pi)$ & $(a^2+b^2+e^2+f^2)^2+2|ac^*+bd^*+eg^*+fh^*|^2+(c^2+d^2+g^2+h^2)^2$ \\
        \eqref{eq:constraint3} & $Y(\pi)$ & $\big[\sqrt{(a^2+h^2)(d^2+e^2)}+\sqrt{(b^2+g^2)(c^2+f^2)}\big]^2$ \\
         & $L(\pi)$ & $(a^2+c^2+f^2+h^2)^2+(b^2+d^2+e^2+g^2)^2$ \\
        \hline
    \end{tabular}
    \caption{Expressions for $X(\pi)$, $Y(\pi)$, and $L(\pi)$ for linear entropy $\EE$ and entropy of coherence $\CC$.}
    \label{tab:xyl-lelc}
\end{table}

We now consider the case where $(\EE,\CC)$ are linear entropy
$\EE^{(C)}_{A,B}\equiv\EE_\ell(|\psi\rangle^{(C)}_{AB})=1-\mathrm{Tr}({\rho^{(C)~2}_{A}})$   
and $\ell^2$-norm of coherence $\CC^{(C)}_A\equiv\CC_{\ell^2}(\rho^{(C)}_A)=\sum_{i\neq j}|\rho^{(C)}_{A,ij}|^2$ respectively, maintaining the assignment procedure from Section \ref{ssec:qrfs}. Let us again begin with the 3-qubit pure state $\ket{\psi}_{ABC}=a\ket{000}+b\ket{001}+c\ket{010}+d\ket{011}+e\ket{100}+f\ket{101}+g\ket{110}+h\ket{111}$, with normalization $a^2+b^2+c^2+\cdots+h^2=1$. Eqs. \eqref{eq:derivation1-3e}-\eqref{eq:derivation1-2c} become
\begin{align}
    &\EE_{\gamma,\alpha\beta}=1-X(\pi), \label{eq:derivation2-3e} \\
    &\EE^{(\alpha)}_{\beta,\gamma}=1-L(\pi)-2Y(\pi), \label{eq:derivation2-2e} \\
    &\CC^{(\alpha)}_\beta=2Y(\pi). \label{eq:derivation2-2c}
\end{align}
where $\alpha$, $\beta$, $\gamma$ encapsulate all permutations of $A,B,C$ and $X(\pi)$, $Y(\pi)$, and $L(\pi)$ are functions of $a,b,c,\dots,h$ that depend on the permutation, $\pi$. The exact expressions for $X(\pi)$, $Y(\pi)$, and $L(\pi)$ for each permutation are listed in Table \ref{tab:xyl-lelc}.

Adding Eqs. \eqref{eq:derivation2-2e} and \eqref{eq:derivation2-2c} and comparing the sum to Eq. \eqref{eq:derivation2-3e}, we obtain equality when
\begin{align} \label{eq:cond2}
    L(\pi)=X(\pi).
\end{align}

It can be shown straightforwardly that the parity states defined in Eqs. \eqref{eq:even} and \eqref{eq:odd} satisfy Eq. \eqref{eq:cond2} for all three constraints. Hence, the parity states respect entanglement transference for linear entropy and $\ell^2$-norm of coherence as well.

\begin{table*}[h]
    \centering
    \begin{tabular}{|c|c|c|}
        \hline
        Quantity & Expression & Scope \\
        \hline
        $\EE^{(A)}_{R,\bar R}$ & $\frac{1}{8}\sin^2(2r)$ & \\
        $\EE^{(R)}_{A,\bar R}$ & $\frac{\sin^2r}{2}$ & Perspectival \\
        $\EE^{(\bar R)}_{A,R}$ & $\frac{\cos^2r}{2}$ & \\
        \hline
        $\EE_{\bar R,AR}$ & $\frac{\sin^2r}{2}\rr{1+\cos^2r}$ & \\
        $\EE_{R,A\bar R}$ & $\cos^2r\rr{1-\frac{\cos^2r}{2}}$ & Global \\
        $\EE_{A,R\bar R}$ & $\frac{1}{2}$ & \\
        \hline
    \end{tabular}
    \caption{Entanglement characterization of the state $\ket{\phi_r}_{AR\bar R}$, where $\EE$ is linear entropy.}
    \label{tab:entanglement2}
\end{table*}

We now turn our attention to the problem of entanglement degradation, specifically, to the state $\ket{\phi_r}_{AR\bar R}$ in Eq. \eqref{eq:global-state}. Since $\ket{\phi_r}_{AR\bar R}$ is a global even state, entanglement transference tells us that we may fully characterize the $\EE+\CC$ properties of the system by providing six entanglement quantities (in this case, the linear entropy of entanglement). We do so in Table \ref{tab:entanglement2}.

\begin{figure*}[h]
    \centering
    \includegraphics[width=0.32\linewidth]{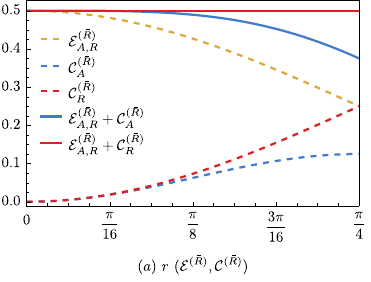}
    \includegraphics[width=0.32\linewidth]{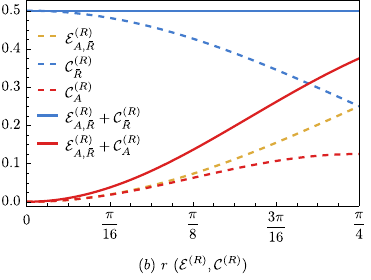}
    \includegraphics[width=0.32\linewidth]{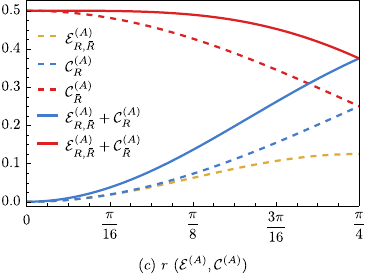}
    \caption{Comparison of linear entropy $\EE$ plus subsystem $\ell^2$-norm coherence $\CC$ from the perspectives of (a) anti-Rob, (b) Rob, and (c) Alice in the three-qubit system. Linear entropy from each perspective is depicted by the yellow dashed line. Blue colour corresponds to the clockwise cyclic direction, defined to be $A\rightarrow R\rightarrow\bar R\rightarrow A$, while red colour corresponds to the anti-clockwise cyclic direction, $A\rightarrow\bar R\rightarrow R\rightarrow A$. For example, in plot (a), the subsystem coherence of Alice from anti-Rob's perspective is given by the blue dashed line, since Alice is cyclically ``downstream'' of anti-Rob. If anti-Rob makes reference to Rob, who is in the reverse direction, then the corresponding subsystem coherence is the red dashed line. The totals $\EE+\CC$ are given in each plot by the solid lines, which follow the same colour convention. Observe that Eqs. \eqref{eq:pair1}-\eqref{eq:pair3} are satisfied, since red solid lines 
    in a given panel are the same as
    the blue solid lines in the (cyclically) right adjacent panel.}
    \label{fig:el-and-cl}
\end{figure*}

Fig. \ref{fig:el-and-cl} shows the linear entropy $\EE$ and subsystem $\ell^2$-norm coherence $\CC$ of the $|\phi_r\rangle_{AR\bar R}$ system for (a) Alice and Rob's qubits from the perspective of anti-Rob, (b) Alice and anti-Rob's qubits from the perspective of Rob, and (c) Rob and anti-Rob's qubits from the perspective of Alice, as a function of the acceleration parameter $r$. As before, Eq. \eqref{eq:pair1}-\eqref{eq:pair3} are satisfied. Entanglement degradation between Alice and Rob is also captured in Fig. \ref{fig:el-and-cl}a. As before, we may find a perspectival coherence quantity, $\CC^{(\bar R)}_R$, that completely offsets the degradation of $\EE^{(\bar R)}_{A,R}$. Their sum, which is equal to the global entanglement $\EE_{A,R\bar R}$ by entanglement transference, is $\frac{1}{2}$ for all values of $r$. We emphasize that the curves in Fig. \ref{fig:el-and-cl} are quantitatively different from those in Fig.~\ref{fig:e-and-c}, despite their qualitative similarities.

\section{MUTUAL INFORMATION} \label{sec:app-mutual-info}

In this Appendix, we comment on mutual information and provide a correction to Fig. 5 in the original paper on entanglement degradation \cite{Alsing2006}.

Mutual information \cite{Ingarden:1997} is a measure of total correlation between two systems, defined as
\begin{align}
    \mathcal{I}=\mathcal{S}(\rho_a)+\mathcal{S}(\rho_b)-\mathcal{S}(\rho_{ab}),
\end{align}
where $a$ and $b$ denote the systems and $\mathcal{S}$ is the von Newmann entropy. Mutual information quantifies how much two correlated observers, one with access to system $a$ and the other with access to system $b$, know about the other observer's state. Note that if $\rho_{ab}$, the joint state of $a$ and $b$ after tracing over all other systems (hypothetically, $c,d,\dots$), is pure, then $\mathcal{S}(\rho_{ab})=0$ and $\mathcal{S}(\rho_a)=\mathcal{S}(\rho_b)$. Hence, mutual information reduces to twice the entanglement entropy of the joint state $\rho_{ab}$. Clearly, the perspectival states $\ket{\psi}^{(A)}_{R\bar R}$, $\ket{\psi}^{(R)}_{A\bar R}$, and $\ket{\psi}^{(\bar R)}_{AR}$ are pure, and hence the expressions for their respective mutual information can be obtained from Table \ref{tab:entanglement1} in Section \ref{sec:qrf-deg} by multiplying a factor of 2. In the global approach, $\rho_{ab}$ will, in general, be mixed, as we must trace over the third system. The analytic expressions for the three non-perspectival cases can also be obtained from Table \ref{tab:entanglement1} as follows:
\begin{align}
    &\mathcal{I}_{R,\bar R}=\EE_{\bar R,AR}+\EE_{R,A\bar R}-\EE_{A,R\bar R}, \\
    &\mathcal{I}_{A,\bar R}=\EE_{A,R\bar R}+\EE_{\bar R,AR}-\EE_{R,A\bar R}, \\
    &\mathcal{I}_{A,R}=\EE_{R,A\bar R}+\EE_{A,R\bar R}-\EE_{\bar R,AR}.
\end{align}

\begin{figure}[h]
    \centering
    \includegraphics[width=0.5\linewidth]{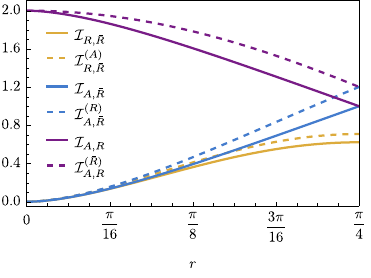}
    \caption{Mutual information between each pair of subsystems in the non-perspectival approach (solid lines) and the perspectival approach (dashed lines).}
    \label{fig:mutual-info}
\end{figure}

Fig. \ref{fig:mutual-info} shows $\mathcal{I}_{R,\bar R}$, $\mathcal{I}_{A,\bar R}$, and $\mathcal{I}_{A,R}$ as a function of the acceleration parameter $r$. The corresponding mutual information in each perspective is also depicted, using dotted lines. First, we observe that the solid curves in Fig. \ref{fig:mutual-info} differ from Fig. 5 in the paper \cite{Alsing2006}. Although the analytic expressions derived in \cite{Alsing2006} are correct, the corresponding plot was inaccurate, and we provide the correction here. Qualitatively, however, a number of similarities remain. The non-perspectival mutual information $\mathcal{I}_{A,R}$ between Alice's and Rob's systems attains a maximum of 2 at $r=0$ and decreases monotonically with acceleration. The mutual information $\mathcal{I}_{A,\bar R}$ between Alice's and anti-Rob's systems increases monotonically with acceleration, symmetric with $\mathcal{I}_{A,R}$ via reflection through the horizontal line $\mathcal{I}=1$. At infinite acceleration, $r=\frac{\pi}{4}$, these two curves intersect at 1. $\mathcal{I}_{R,\bar R}$ between Rob and anti-Rob also increases monotonically, plateauing at $\mathcal{I}=\frac{3}{2}\rr{2-\log_2 3}$.

We observe that the mutual information between any pair of subsystems in the non-perspectival approach is \emph{not} the same as the mutual information between the same pair of subsystems, but viewed from the perspective of the third subsystem (i.e., the solid lines do not coincide with the dashed lines). For example, the purple dotted line, which represents the mutual information between Alice's and Rob's systems from the perspective of anti-Rob, degrades more slowly than the purple solid line, which is the usual mutual information between Alice's and Rob's systems. Thus, the intersection of $\mathcal{I}^{(\bar R)}_{A,R}$ and $\mathcal{I}^{(R)}_{A,\bar R}$ occurs asymmetrically at $-(1+\frac{1}{\sqrt{2}})\log_2(\frac{1}{2}+\frac{1}{2\sqrt{2}})-(1-\frac{1}{\sqrt{2}})\log_2(\frac{1}{2}-\frac{1}{2\sqrt{2}})\approx1.20>1$. One sees in Fig. \ref{fig:mutual-info} that all perspectival (dashed line) quantities are greater than their non-perspectival (solid line) counterparts, suggesting that the perspectival approach may be correlation preserving.

\end{widetext}
\end{document}